**Penetration Testing and Legacy Systems**

Sandra P. Smyth

Purdue University

**Author Note**

Sandra P. Smyth PUID:   0033637631

Email: smyth4@purdue.edu

**Abstract**

As per Adusumilli (2015), "70% of corporate business systems today are legacy applications. Recent statistics prove that over 60% of IT budget is spent on maintaining these Legacy systems, showing the rigidity and the fragile nature of these systems."

Usually, testing is included during the software development cycle, using testing techniques such as unit testing, integration testing, and system testing before releasing the product. After the software product is released to production, no additional testing is done; the testing process is back to the table only when modifications are made. Techniques such as regression testing are included to ensure the changes do not affect existing functionality, but testing non-functional features that are rarely included in such regression tests' scope.

Schrader (2021) affirms that "legacy systems are often maintained only to ensure function," and IT organizations may fail to consider the cybersecurity perspective to remain secure. Legacy systems are a high-risk component for the organization that must be carefully considered when structuring a cyber security strategy. This paper aims to help the reader understand some measures that can be taken to secure legacy systems, explaining what penetration testing is and how this testing technique can help secure legacy systems.

## Penetration Testing and Legacy Systems

### 1.        The Problem

As per Adusumilli (2015), "70% of corporate business systems today are legacy applications." And as Schrader (2021) affirms, "legacy systems are often maintained only to ensure function." Thus, legacy systems become a high cyber security risk making the organization are an easy target.   Since those systems are not supported or updated, their vulnerabilities are kept as an open invitation for the hackers. This paper aims to help the reader understand some measures that can be taken to secure legacy systems, explaining what penetration testing is and how this testing technique can help secure legacy systems.

### 2.        What is penetration testing?

Penetration testing is also known as **Pentesting** or **Pen Test**.  **Pentester, Vulnerability Assessor, White Hat Hacker, or Ethical Hacker** are different names to call the cyber security expert who executes this type of testing.

As NetSPI (n.d.) defines it, "Penetration testing simulates the actions of a skilled threat actor determined to gain privileged access." Its main goal is to identify vulnerabilities in high-risk IT systems, revealing security weaknesses.  The following are some examples of the types of vulnerabilities that penetration testing can reveal as per (Bit Sentinel, n.d.)

- "Weaknesses in your infrastructure setup
- Flaws in the operating systems, services, and applications used throughout your company
- Improper configurations
- Risky end-user behavior
- Logic flaws in the applications' business processes
- Weak credentials that cybercriminals can use for malicious purposes
- Hijacking"

It is important to differentiate between penetration testing and vulnerability scanning to know that these are two different practices.  As per ControlScan (2017), "Vulnerability scans look for known vulnerabilities in your systems and report potential exposures.  Penetration tests are intended to exploit weaknesses in the architecture of your IT network and determine the degree to which a malicious attacker can gain unauthorized access to your assets.  A vulnerability scan is typically automated, while a penetration test is a manual test performed by a security professional." According to CoreSecurity on the 2021 Pen Testing Survey Report, 10% of the companies responded that they perform daily penetration testing; CoreSecurity believes they are more probably be executing automated vulnerability testing and not PenTest.

### I.    The benefits

Figure 1 lists the benefits of the Pen Test according to NetSPI.

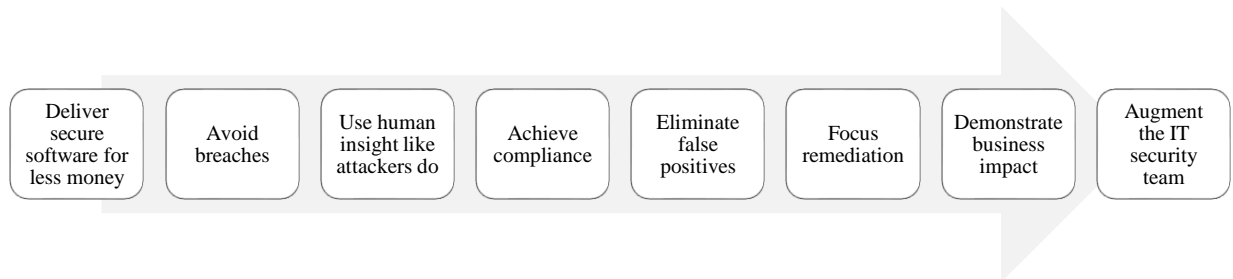

*Figure 1.  Benefits of penetration testing (Smyth, 2022)*

NetSPI mentioned that Pen Test helps harden the environment proactively, which is the main benefit of this practice.

### II.    The drawbacks

Perez Araya (2018) suggests the following potential drawbacks from PenTest:

- "Outages to critical services if the pen test is poorly designed or executed, which can end up causing more damage to the company in general; and

- Difficulty conducting pen tests on legacy systems, which are often vital to businesses."

Spencer (2020) brings up an interesting point, referring to the fact that PenTest tools are used at the end of the SDLC, and "the time required to triage, diagnose, and remediate the potential vulnerabilities that are found often spirals out of control.  Research shows that vulnerabilities found in testing cost 10x more to fix than those detected in coding.  Waiting until the beta is 19x more and until the actual release is 29x more."

Another challenge highlighted by CoreSecurity (2021) is that "Running a penetration test exposes security weaknesses, raising awareness about potential attack vectors.  But if you don't fix the problems that you uncover and a breach occurs, you risk culpability for not acting on an issue that you knew existed." See figure 2 for the Pentesting challenges presented by CoreSecurity (2021).

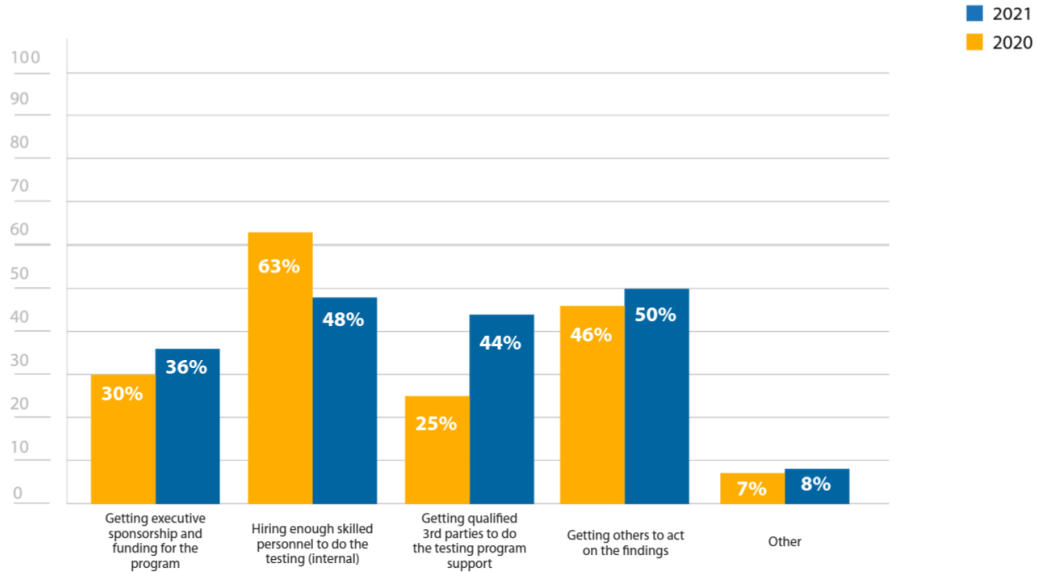

*Figure 2.  What challenge(s) does your organization face with your penetration testing program?*
*(CoreSecurity, 2021)*

## III.    Types of Pentesting

Figure 3 shows the different types of Pentesting presented by NetSPI (n.d.), which can target networks, software applications, the cloud, and employee behaviors.  The focus of this paper is deep into the application penetration testing type for legacy systems.

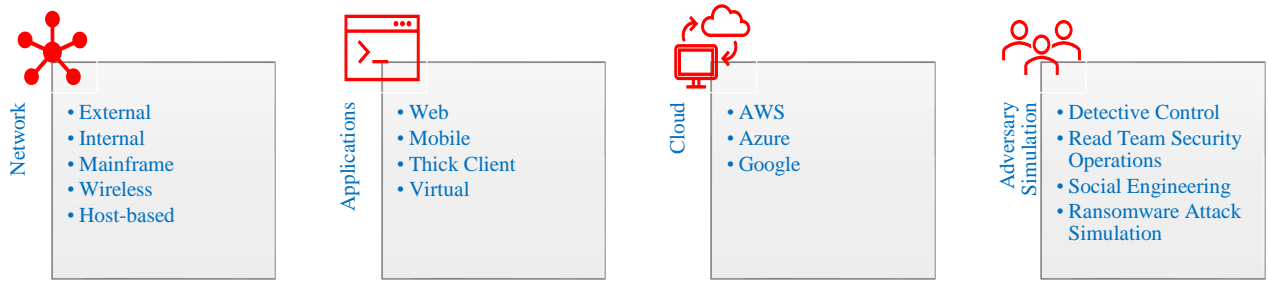

*Figure 3.  Types of penetration testing (Smyth, 2022).*

## 3.    What is a legacy system?

As Gartner's (n.d.) Information Technology glossary describes it as "An information system that may be based on outdated technologies, but is critical to day-to-day operations." Techopedia (2014) clarifies that age does not define legacy systems.  Figure 4 lists some characteristics that help to identify a legacy system.

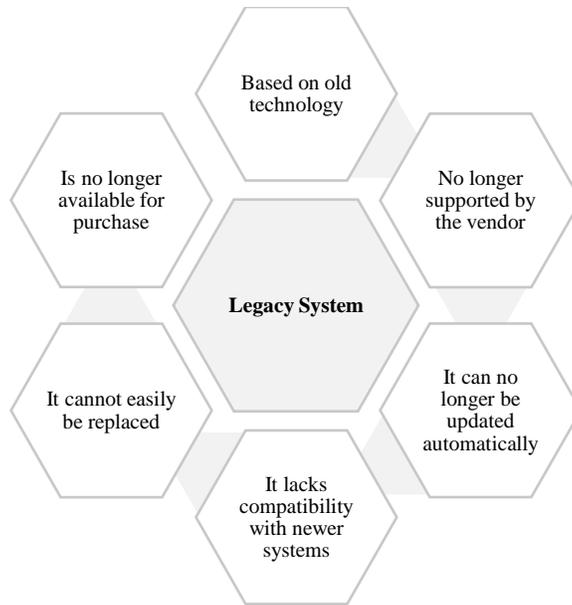

*Figure 4.  Legacy Systems - Common characteristics (Smyth, 2022)*

Brehm (2021) provides examples of legacy software still running in government organizations and private companies.  See figure 5.

Windows 7 officially became a legacy operating system in January 2020 after Microsoft halted security updates and support for it. However, over 100 million machines continue to run this operating system.

Common Business-Oriented Language or COBOL is still used 55 years after its development. Forty-eight percent of businesses and government organizations reportedly depend on this language more than others.

Discontinued Oracle products: Oracle database software such as E-Business Suite and Peoplesoft

*Figure 5.  Examples of legacy software (Smyth, 2022).*

The use of legacy systems can introduce different risks and issues to the security organization landscape; Talend (n.d.) has an excellent way to synthesize it, "As technology advances, risks increase for legacy systems."  Some of the issues that the use of legacy systems presents are: isolating data from other systems due to the lack of compatibility; high maintenance costs due to the lack of support from the official vendor; slow performance, and productivity due to the lack of application updates.  The risk that this paper explores is the security breaches that legacy system adds to the IT environment due to the vendor's lack of support, updates, and maintenance and the use of old protocols and standards.

### 4.        Legacy Systems' security vulnerabilities

Vulnerabilities on legacy systems come in different flavors, as listed in figure 4, leaving holes in an organization's security hygiene, as presented by IDI Billing (2019).  "Cybersecurity best practices have evolved significantly over the last several years, and legacy systems may be ill-equipped to keep up with those developments.  When you consider how much of a focal point IT security has become in the wake of recent data breaches and scandals, legacy systems often fall woefully short of modern benchmarks." (para.14)

Armerding (2020) points out that "Any *'legacy code'* that hasn't been reviewed recently, no matter how well it appears to work, might contain vulnerabilities."  Some possible scenarios are:

- The code was obtained as an open-source solution, and the creator of that code is not providing any more updates.
- The code is working as expected but has not been tested using current testing procedures that include security test cases.
- Another case could be those actively developed solutions for which new features and functionalities are added, but the testing procedures are only applied to the newly added code instead of executing regression testing that includes current testing practices.

InfoSecurity (2019) warns that "this continued use of legacy systems presents a major security risk as developers focus on actively providing support for their latest versions.  So, any discovered or disclosed vulnerability for these older systems may not be fixed or addressed, leaving them vulnerable to attack."

As previously mentioned, legacy systems lack compatibility characteristics.  They are difficult, too costly, or impossible to integrate with new technologies, making them the weakest link in the cyber security chain.  Synchrony Systems (2020) provides examples when it states, "Legacy systems may be incompatible with security features surrounding access, such as multi-factor authentication, single-sign-on, and role-based access, or lack sufficient audit trails or encryption methods.  Whatever the reason, these systems are unable to accommodate today's security best practices."

The fact that a system is working does not guarantee that they are safe.  Some companies prefer not to patch the old systems to avoid unwanted modifications to their functionality.  InfoSecurity (2019) mentioned this lack of interest from companies as a cause for keeping alive vulnerabilities that have been already resolved, and it shows us that those are not isolated cases, "An RSA Conference survey revealed that less than half of companies patch vulnerabilities once they are publicized.  Some even wait weeks or months before acting on security bulletins." (para.9)

According to Automox (n.d.), "Many legacy systems are still supported, and it's important to regularly patch them.  Estimates suggest unpatched vulnerabilities are linked to 60 percent of data breaches – and further studies have found that some 44 percent of exploits target a vulnerability that is two to four years old.  This means most data breaches come from unpatched vulnerabilities – and in many cases, attackers are exploiting vulnerabilities that were discovered years ago and can be used to target legacy systems."

## 5.  Basic security measurements to protect Legacy Systems

Armerding (2020) highlights the importance of keeping legacy systems 'current'; the need to evaluate, find and fix vulnerabilities as soon as possible.  The recommendation from Weber (2006) is to assess each legacy system individually, and the security risk that each system faces and then consider mitigation options accordingly.  See figure 6.

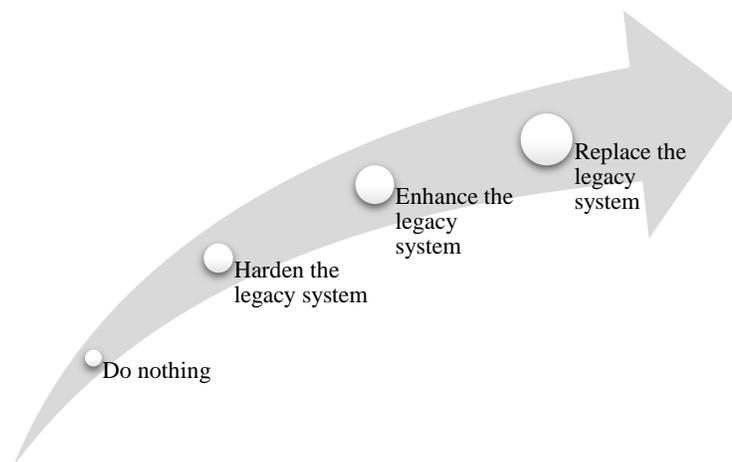

*Figure 6.  Legacy Systems - Mitigation options (Smyth, 2022)*

Figure 7 shows the to-do list proposed by InfoSecurity (2019), with some basic measures to protect the legacy system.

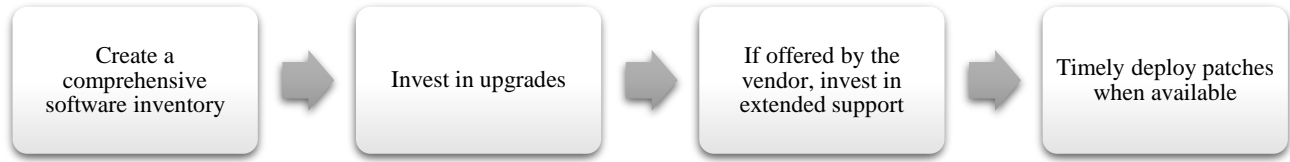

*Figure 7.  Recommendation to protect legacy systems (Smyth, 2022)*

## 6.        How is penetration testing done?

Figure 8 shows the phases of penetration testing according to NetSPI (n.d.).  During the planning phase, the pentester defines the scope and objectives of the test.  In the second phase and by using scanning tools, vulnerabilities are identified.  And to avoid false positives, NetSPI recommends only enumerating those vulnerabilities that are identified by multiple tools.  The pentester uses penetration testing tools and manual techniques in the third phase to evaluate the found vulnerabilities to identify and validate exploitable entry points.  During phase fourth, the goal is to gain privileged access status on a networked device and jump between trusted networked zones.  The vulnerabilities that allow those exploits should be adequately documented during this phase.  Lastly, the findings need to be ranked by severity and reported.  As NetSPI indicates, "The report includes step-by-step instructions for how to reproduce the pentester's exploits so remediators can reproduce and fix them."

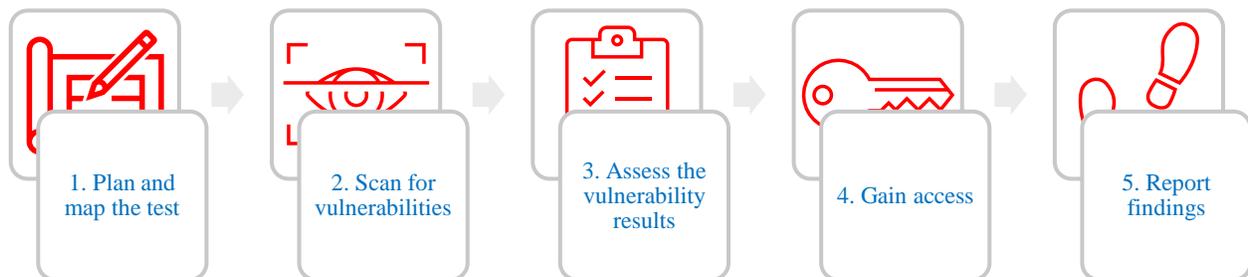

*Figure 8.  Phases of penetration testing (Smyth, 2022)*

NetSPI recommends executing the pentesting on a sandbox environment that simulates the production environment, that way, the testing can be more rigorous.  When testing is done in production systems, the test cases may be limited to read-only scenarios.

## I.    Methodologies and frameworks

Besides the phases described above, several published penetration testing methodologies and frameworks can be used to conduct a more structured penetration test.  The following are five penetration testing methodologies and frameworks that companies use:

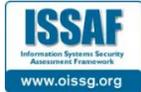 **Information System Security Assessment Framework (ISSAF).** (Chebbi, 2018)

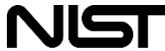 **Technical Guide to Information Security Testing and Assessment (NIST).** (Scarfone et al., 2008)

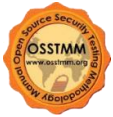 **Open-Source Security Testing Methodology Manual (OSSTMM).** (ISECOM, n.d.)

**Open Web Application Security Project (OWASP).** (OWASP Foundation, n.d.)

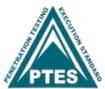 **Penetration Testing Methodologies and Standards (PTES).** (PTES, n.d.)

## II.   Tools

According to NetSPI (n.d.), test cases for common exploits can be automated by using automated penetration testing tools.  Also, NetSPI listed the most popular penetration testing tools as follows:

- "**PowerUpSQL**: a PowerShell toolkit for attacking SQL Server

- **SQL Injection Wiki**: a wiki focused on aggregating and documenting SQL injection methods

- **MicroBurst**: a collection of scripts for assessing Microsoft Azure security

- **PESecurity**: a PowerShell module to check if a Windows binary (EXE/DLL) was compiled with security features such as ASLR (Address Space Layout Randomization), DEP (Data Execution Prevention), and SafeSEH (Structured Exception Handling), Strong Naming, and Authenticode

- **Goddi** (go dump domain info): a tool that dumps Active Directory domain information."

## III.   How often should a PenTest be scheduled?

According to CoreSecurity (2021), there is not a magic formula to determine the frequency on which penetration testing should be performed, but variables like the size of the company, the resources to be used, the scope of the test, and compliance with regulatory laws should be considered when making that decision.  "For example, if a business accepts credit cards, they must comply with the Payment Card Industry Data Security Standard (PCI DSS) 3.2.1.  Requirement 11 of the PCI DSS states that "system components, processes, and custom software should be tested frequently to ensure security controls continue to reflect a changing environment." The Standard requires that penetration testing should be performed at least annually or whenever there is a significant upgrade or modification of the infrastructure and applications in use."(RSI Security, 2020)

See figure 9 provided by CoreSecurity (2021) with the reasons organizations gave for not using pentesting.

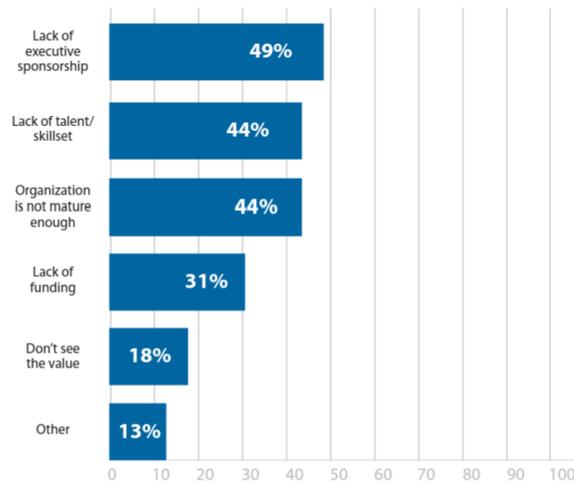

*Figure 9.  Why does your organization not conduct penetration tests? (CoreSecurity, 2021)*

On the other hand, Bit Sentinel (2019) suggest performing penetration testing "on a regular basis (at least once a year) to ensure more consistent IT and network security management by revealing how newly discovered threats (0-days, 1-days) or emerging vulnerabilities might be exploited by malicious hackers." Bit Sentinel also adds that in addition to the scheduled PenTest, additional runs should be planned whenever:

- "New network infrastructure or applications are added
- Significant upgrades or modifications are applied to infrastructure or applications
- New office locations are established
- Security patches are applied
- End-user policies are modified." Bit Sentinel (2019)

## Conclusion

It is well known that legacy systems represent a high risk for the organization's security. Willis (2019) recognizes that those systems are still critical and that many companies hesitate to deploy patches or update such systems because they are afraid of malfunctions. However, Willis's response to this concern is as follows: "A malfunction with an update might cause some downtime – but a data breach would be a much bigger problem. If the current infrastructure is too old to allow patching, it must be updated; legacy systems shouldn't be used as an excuse not to patch. If they're unsupported, then retire them; if they are still supported, patch them!"

It is recommended to use pentesting to identify vulnerabilities and document step-by-step to reproduce them so that they can be fixed. The type of found vulnerabilities will depend on the outlined plan for the test; for example, the scope of the test can be to identify flaws within the security policy. As per Spencer, 2020, "70% of CIOs report their teams spend more than half of their time finding the cause before they're able to fix system problems." Thus, the inclusion of a well PenTest plan can help reduce the time spent on finding the causes of problems.

Nearly all companies will agree that testing is important, however, and as Perez Araya (2018) states, "Most companies do not adhere to this recommendation because they are eager to get their return on investment (ROI) quickly. Companies might also fail to follow this best practice because a project has exceeded its deadline or budget. These factors make companies enthusiastic to push their new services live without having conducted the proper security assessments. This is a risk that needs to be evaluated and put in perspective when deploying new systems." Also valid, for maintaining current the systems that are still operational.

Ruff (2020) summarizes the cost problem of maintaining legacy systems from the financial view: "ROI on implementing software is not always good. The CEOs want to invest on digital transformation but they rather to implement a new system than fix or invest on the current legacy systems."

# References

Adusumilli, S. (2015, April 9). Testing a legacy application with zero documentation. *Software Testing Blog by Cigniti Technologies*. https://www.cigniti.com/blog/testing-a-legacy-application-with-zero-documentation/

Armerding, T. (2020, March 23). *Legacy vulnerabilities: How to find and remediate them*. Software Integrity Blog. https://www.synopsys.com/blogs/software-security/how-to-deal-with-legacy-vulnerabilities/

Automox. (n.d.). *Unpatched vulnerabilities make legacy systems easypPrey*. Retrieved March 27, 2022, from https://www.automox.com/blog/unpatched-vulnerabilities-make-legacy-systems-easy-prey

Bit Sentinel. (n.d.). Certified independent penetration testing. *Bit Sentinel*. Retrieved March 28, 2022, from https://bit-sentinel.com/penetration-testing/

Bit Sentinel. (2019, June 1). How often should I perform penetration testing? *Bit Sentinel*. https://bit-sentinel.com/how-often-should-i-perform-penetration-testing/

Brehm, T. (2021, June 9). *What is a legacy system and legacy software?* Entrance. https://www.entranceconsulting.com/what-is-legacy-system-and-legacy-software/

Chebbi, C. (2018). *Advanced infrastructure penetration testing*. Packt Publishing. https://learning.oreilly.com/library/view/advanced-infrastructure-penetration/9781788624480/

ControlScan. (2017, May 30). *Penetration tests vs. vulnerability scans: How are they different?* ControlScan. https://www.controlscan.com/penetration-tests-vulnerability-scans-different/

CoreSecurity. (2021a). *2021 Pen testing survey report*. https://www.coresecurity.com/resources/guides/2021-pen-testing-survey-report/thank-you?submissionGuid=6c361970-51de-4bcc-bcf2-5f8f78ecc84f

CoreSecurity. (2021b). *Penetration testing frequency: How often should you test?* https://www.coresecurity.com/blog/penetration-testing-frequency-how-often-should-you-pen-test

Gartner. (n.d.). *Definition of legacy application or system*. Gartner. Retrieved March 19, 2022, from https://www.gartner.com/en/information-technology/glossary/legacy-application-or-system

IDI Billing. (2019, November 18). *The security risks of holding onto legacy systems*. IDI Billing. https://www.idibilling.com/resources/blog/what-are-the-security-risks-of-holding-onto-legacy-systems/

InfoSecurity. (2019, April 18). *How forgotten legacy systems could be your downfall*. Infosecurity Magazine. https://www.infosecurity-magazine.com/opinions/forgotten-legacy-downfall-1/

ISECOM. (n.d.). *OSSTMM 3 – The open source security testing methodology manual*. Retrieved March 20, 2022, from https://www.isecom.org/research.html

NetSPI. (n.d.). *Penetration testing*. NetSPI. Retrieved January 25, 2022, from https://www.netspi.com/penetration-testing-security/

OWASP Foundation. (n.d.). *OWASP Web security testing guide*. Retrieved March 20, 2022, from https://owasp.org/www-project-web-security-testing-guide/

Perez Araya, W. (2018, August 21). Why, when and how often should you Pen Test? *Security Intelligence*. https://securityintelligence.com/why-when-and-how-often-should-you-pen-test/

PTES. (n.d.). *The penetration testing execution standard*. Retrieved March 20, 2022, from http://www.pentest-standard.org/index.php/Main_Page

RSI Security. (2020, April 14). How often should you conduct penetration testing? *RSI Security*. https://blog.rsisecurity.com/how-often-should-you-run-penetration-testing/

Ruff, T. (2020, May 14). *When legacy becomes loss: The true costs of legacy systems*. Www.Userlane.Com. https://www.userlane.com/enterprise-legacy-systems/

Scarfone, K., Souppaya, M., Cody, A., & Orebaugh, A. (2008). *Technical guide to information security testing and assessment* (NIST Special Publication (SP) 800-115; p. NIST SP 800-115). National Institute of Standards and Technology. https://doi.org/10.6028/NIST.SP.800-115

Schrader, D. (2021, May 27). *Prevention is the only cure: The dangers of legacy systems*. Dark Reading. https://www.darkreading.com/vulnerabilities---threats/prevention-is-the-only-cure-the-dangers-of-legacy-systems/a/d-id/1341075

Spencer, P. (2020, July 23). When legacy application security becomes your "Mr. Hyde."
        *Security Boulevard*. https://securityboulevard.com/2020/07/when-legacy-application-
        security-becomes-your-mr-hyde/

Synchrony Systems. (2020, February 10). *5 ways legacy systems add to cybersecurity risks*.
        https://sync-sys.com/5-ways-your-legacy-systems-may-add-to-cybersecurity-risks/

Talend. (n.d.). *What is a legacy system?* Talend - A Leader in Data Integration & Data Integrity.
        Retrieved March 19, 2022, from https://www.talend.com/resources/what-is-legacy-
        system/

Weber, C. C. (2006, December 14). *Assessing security risk in legacy systems*.
        https://www.cisa.gov/uscert/bsi/articles/best-practices/legacy-systems/assessing-security-
        risk-in-legacy-systems

Willis, V. (2019, August 1). *Unpatched vulnerabilities make legacy systems easy prey*.
        https://www.automox.com/blog/unpatched-vulnerabilities-make-legacy-systems-easy-
        prey